\documentclass{nature}
\usepackage{textcase}
\usepackage[utf8]{inputenc}

\usepackage{amssymb}
\usepackage{amsmath}
\usepackage{wasysym}
\usepackage{graphicx}
\usepackage[usenames, dvipsnames]{color}
\usepackage[normalem]{ulem}
\usepackage{soul}
\usepackage[utf8]{inputenc}
\usepackage[T1]{fontenc}
\usepackage[flushleft]{threeparttable}

\usepackage[mathlines]{lineno}
\usepackage[numbers, sort, compress]{natbib}

\usepackage{hyperref}

\usepackage{subfig}
\usepackage[font=footnotesize]{caption}

\usepackage{floatrow}
\usepackage[shortcuts]{extdash}

\DeclareFloatFont{tiny}{\scriptsize}
{\bf \title{Orbital Modulation of Gamma-Rays Beyond 100 TeV from LS~5039}}

\begin{document}

\maketitle

\author{
R.~Alfaro$^{1}$,
M.~Araya$^{2}$,
J.C.~Arteaga-Velázquez$^{3}$,
D.~Avila Rojas$^{4}$,
H.A.~Ayala Solares$^{5}$,
R.~Babu$^{6}$,
P.~Bangale$^{7}$,
E.~Belmont-Moreno$^{1}$,
A.~Bernal$^{4}$,
K.S.~Caballero-Mora$^{8}$,
T.~Capistrán$^{9}$,
A.~Carramiñana$^{10}$,
S.~Casanova$^{11}$,
U.~Cotti$^{3}$,
J.~Cotzomi$^{12}$,
S.~Coutiño de León$^{13}$,
D.~Depaoli$^{14}$,
P.~Desiati$^{15}$,
N.~Di Lalla$^{16}$,
R.~Diaz Hernandez$^{10}$,
B.L.~Dingus$^{17,18}$,
M.A.~DuVernois$^{15}$,
K.~Engel$^{18}$,
T.~Ergin$^{6}$,
C.~Espinoza$^{1}$,
K.L.~Fan$^{18}$,
K.~Fang$^{15}$,
J.A.~García-González$^{19}$,
H.~Goksu$^{14}$,
A.~Gonzalez Muñoz$^{1}$,
J.A.~González$^{3}$,
M.M.~González$^{4}$,
J.A.~Goodman$^{18}$,
S.~Groetsch$^{15}$,
J.P.~Harding$^{17}$,
S.~Hernández-Cadena$^{21}$,
I.~Herzog$^{6}$,
D.~Huang$^{18,20}$,
F.~Hueyotl-Zahuantitla$^{8}$,
P.~Hüntemeyer$^{20}$,
S.~Kaufmann$^{22}$,
D.~Kieda$^{23}$,
A.~Lara$^{24}$,
J.~Lee$^{25}$,
H.~León Vargas$^{1}$,
J.T.~Linnemann$^{6}$,
A.L.~Longinotti$^{4}$,
G.~Luis-Raya$^{22}$,
K.~Malone$^{17}$,
O.~Martinez$^{12}$,
J.~Martínez-Castro$^{26}$,
J.A.~Matthews$^{27}$,
P.~Miranda-Romagnoli$^{28}$,
J.A.~Morales-Soto$^{3}$,
E.~Moreno$^{12}$,
M.~Mostafá$^{7}$,
M.~Najafi$^{20}$,
L.~Nellen$^{29}$,
M.U.~Nisa$^{6}$,
N.~Omodei$^{16}$,
E.~Ponce$^{12}$,
Y.~Pérez Araujo$^{1}$,
E.G.~Pérez-Pérez$^{22}$,
C.D.~Rho$^{30}$,
A.~Rodriguez Parra$^{3}$,
D.~Rosa-González$^{10}$,
M.~Roth$^{17}$,
H.~Salazar$^{12}$,
D.~Salazar-Gallegos$^{6}$,
A.~Sandoval$^{1}$,
M.~Schneider$^{18}$,
G.~Schwefer$^{14}$,
J.~Serna-Franco$^{1}$,
A.J.~Smith$^{18}$,
Y.~Son$^{25}$,
R.W.~Springer$^{23}$,
O.~Tibolla$^{22}$,
K.~Tollefson$^{6}$,
I.~Torres$^{10}$,
R.~Torres-Escobedo$^{21}$,
R.~Turner$^{20}$,
E.~Varela$^{12}$,
X.~Wang$^{20}$,
Z.~Wang$^{18}$,
I.J.~Watson$^{25}$,
H.~Wu$^{15}$,
S.~Yu$^{5}$,
S.~Yun-Cárcamo$^{18}$,
H.~Zhou$^{21}$,
C.~de León$^{3}$
}
\begin{affiliations}
\small
  \item Instituto de F\'{i}sica, Universidad Nacional Autónoma de México, Ciudad de Mexico, Mexico
  \item Universidad de Costa Rica, San José 2060, Costa Rica
  \item Universidad Michoacana de San Nicolás de Hidalgo, Morelia, Mexico
  \item Instituto de Astronom\'{i}a, Universidad Nacional Autónoma de México, Ciudad de Mexico, Mexico
  \item Department of Physics, Pennsylvania State University, University Park, PA, USA
  \item Department of Physics and Astronomy, Michigan State University, East Lansing, MI, USA 
  \item Temple University, Department of Physics, 1925 N. 12th Street, Philadelphia, PA 19122, USA
  \item Universidad Autónoma de Chiapas, Tuxtla Gutiérrez, Chiapas, México
  \item Università degli Studi di Torino, I-10125 Torino, Italy
  \item Instituto Nacional de Astrof\'{i}sica, Óptica y Electrónica, Puebla, Mexico
  \item Institute of Nuclear Physics Polish Academy of Sciences, PL-31342 IFJ-PAN, Krakow, Poland
  \item Facultad de Ciencias F\'{i}sico Matemáticas, Benemérita Universidad Autónoma de Puebla, Puebla, Mexico
  \item Instituto de Física Corpuscular, CSIC, Universitat de València, E-46980, Paterna, Valencia, Spain
  \item Max-Planck Institute for Nuclear Physics, 69117 Heidelberg, Germany
  \item Department of Physics, University of Wisconsin-Madison, Madison, WI, USA
  \item Department of Physics, Stanford University, Stanford, CA 94305–4060, USA
  \item Los Alamos National Laboratory, Los Alamos, NM, USA
  \item Department of Physics, University of Maryland, College Park, MD, USA
  \item Tecnologico de Monterrey, Escuela de Ingenier\'{i}a y Ciencias, Ave. Eugenio Garza Sada 2501, Monterrey, N.L., Mexico, 64849
  \item Department of Physics, Michigan Technological University, Houghton, MI, USA
  \item Tsung-Dao Lee Institute, Shanghai Jiao Tong University, Shanghai, China
  \item Universidad Politecnica de Pachuca, Pachuca, Hgo, Mexico
  \item Department of Physics and Astronomy, University of Utah, Salt Lake City, UT, USA
  \item Instituto de Geof\'{i}sica, Universidad Nacional Autónoma de México, Ciudad de Mexico, Mexico
  \item University of Seoul, Seoul, South Korea
  \item Centro de Investigaci\'on en Computaci\'on, Instituto Polit\'ecnico Nacional, M\'exico City, M\'exico
  \item Dept of Physics and Astronomy, University of New Mexico, Albuquerque, NM, USA
  \item Universidad Autónoma del Estado de Hidalgo, Pachuca, Mexico
  \item Instituto de Ciencias Nucleares, Universidad Nacional Autónoma de Mexico, Ciudad de Mexico, Mexico
  \item Department of Physics, Sungkyunkwan University, Suwon 16419, South Korea
\end{affiliations}
\noindent {\bf Gamma-ray binaries are systems composed of a compact object orbiting a massive companion star. The interaction between these two objects can drive relativistic outflows, either jets or winds, in which particles can be accelerated to energies reaching hundreds of tera-electronvolts (TeV) \cite{1998A&A...338L..71M,2002A&A...384..954R,2002A&A...393L..99P,dubus}. It is however still debated where and under which physical conditions particles are accelerated in these objects and ultimately whether protons can be accelerated up to PeV energies. Among the well-known gamma-ray binaries, LS~5039 is a high-mass X-ray binary (HMXB) with an orbital period of 3.9 days that has been observed up to TeV energies by the High Energy Stereoscopic System (H.E.S.S.) \cite{hessls5039,2005Sci...309..746A}. 
In this work, we present new observations of LS~5039 obtained with the High Altitude Water Cherenkov (HAWC) observatory. Our data reveal that the gamma-ray spectrum of LS~5039 extends up to 200~TeV with no apparent spectral cut-off. 
Furthermore, we confirm, with a confidence level of 4.7$\sigma$, that the emission between 2 TeV and 118 TeV is modulated by the orbital motion of the system, which indicates that these photons are likely produced within or near the binary orbit where they can undergo absorption by the stellar photons. In a leptonic scenario, the highest energy photons detected by HAWC can be emitted by $\sim200$ TeV electrons inverse Compton scattering stellar photons, which would require an extremely efficient acceleration mechanism operating within LS~5039. Alternatively, a hadronic scenario could explain the data through proton-proton or proton-$\gamma$ collisions of protons accelerated to peta-electronvolt (PeV) energies.}

 LS 5039, located at a distance of $\sim3$~kpc from the Sun \cite{1997A&A...323..853M}, is a well-known binary system that was first discovered through a Roentgensatellit~(ROSAT) survey in 1997 \cite{1997A&A...323..853M,1998A&A...338L..71M}. 
LS~5039 consists of a compact object in orbit around an O6.5V type star \cite{2000Sci...288.2340P}.
The average binary orbital radius is ($1.4-2.9$) $\times 10^{12}$ cm \cite{2005MNRAS.364..899C} with an orbital period of $\sim3.9$~days \cite{Bosch-Ramon_2005}. 

LS~5039 has been long thought to be a microquasar candidate \cite{1999ARA&A..37..409M}. However, a lack of accretion signatures in X-ray observations together with recent evidence of 9s long X-ray pulsations from the compact object \cite{PhysRevLett.125.111103} have led many to support the pulsar binary model as well. Although LS~5039 has been observed and studied across a wide range of the electromagnetic spectrum since its discovery, we do not fully understand its properties and the underlying physical mechanisms governing the source. A follow-up analysis at multi-TeV energies can help shed light on the acceleration mechanisms and radiation processes within LS~5039.

\begin{figure}[ht!]
\begin{center}
\resizebox{0.9\textwidth}{!}{%
\includegraphics[width=0.5\textwidth]{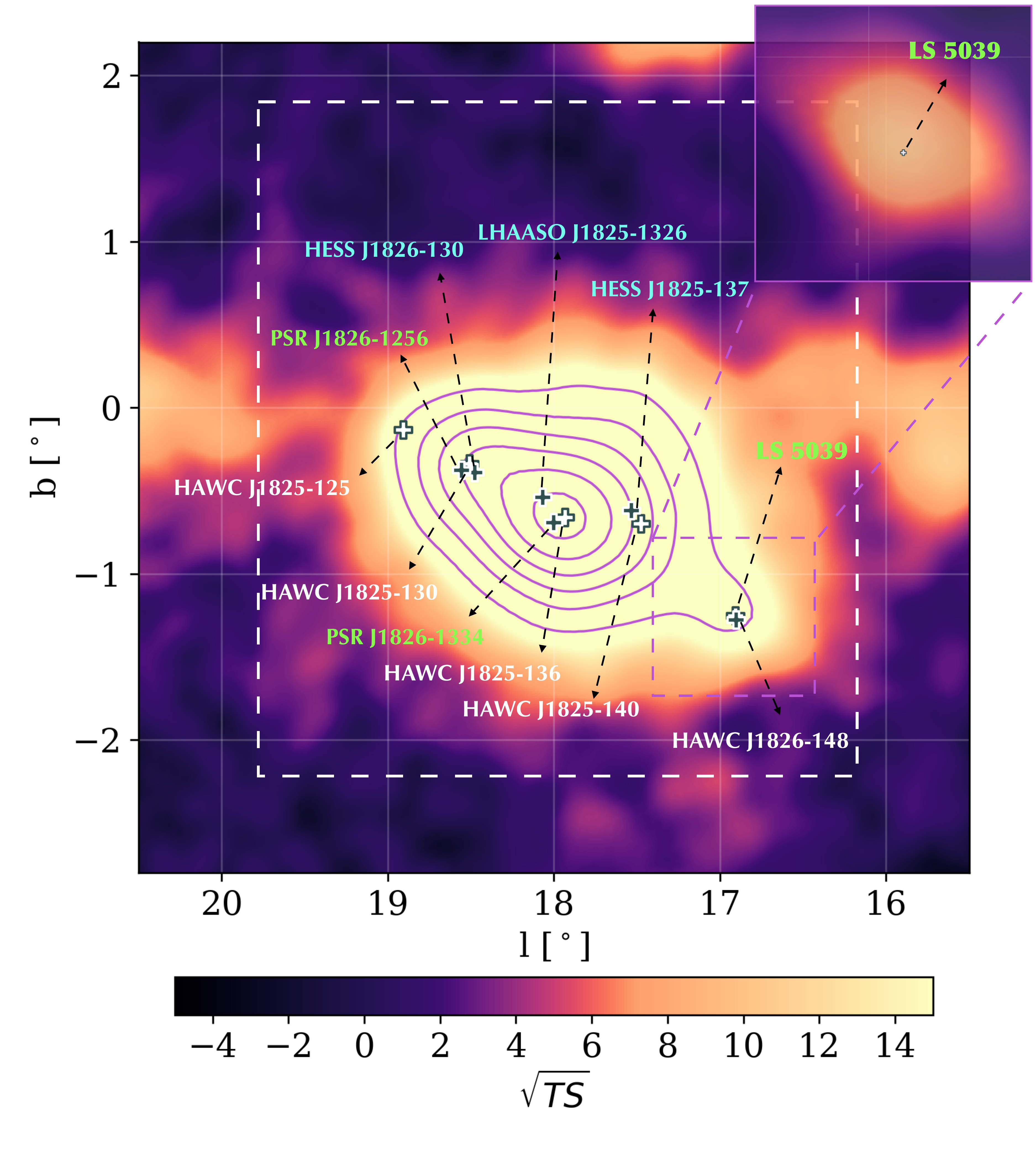}%
}
\caption{HAWC significance map of the LS5039 region. The dashed box indicates the ROI for this work. White labels mark sources resolved in this analysis; light green labels indicate the known astrophysical sources inside the ROI, with LS~5039 highlighted in bold text; cyan labels show measurements from other observatories. Significance contours start at $20\sigma$, increasing by $10\sigma$ per line. Upper Right: Zoomed-in HAWC significance map around LS5039 after background sources subtraction (see Methods~\ref{tab:modellist}), showing an $11\sigma$ hotspot at LS~5039's known location.}
    
\label{fig:map}
\end{center}
\end{figure}

The HAWC gamma-ray observatory is a ground-based particle sampling array with an instantaneous 2~sr field-of-view and $>95$\% duty cycle, which makes it suitable for continuous observations of different orbital phases of binary systems. Figure~\ref{fig:map} shows the significance map of the complex region around LS~5039 produced from 2,859 days of cumulative HAWC data. The application of a source search pipeline (see Methods~\ref{sec:sss}) allows us to deconvolve the region of interest (ROI), indicated by the white dashed rectangular box, and separate the observed gamma-ray emission from individual sources. The top right inset in Figure~\ref{fig:map} shows a zoomed in map of the LS~5039 excess after subtracting the identified background sources.
All the known TeV gamma-ray sources have been labeled on this plot, which includes LHAASO~J1825-1326 \cite{cao2021ultrahigh} observed by the Large High Altitude Air Shower Observatory (LHAASO), and HESS~J1825-137 and HESS~J1826-130 \cite{abdalla2018hess} observed by H.E.S.S.. We have separately indicated sources that have prefixes, ``HAWC'', which are the background sources identified in this work. While the HAWC sources (HAWC~J1825-130, HAWC~J1825-136, and HAWC~J1825-140) have good spatial agreement with the sources from H.E.S.S. and LHAASO, we have also identified one extra source, HAWC~J1825-125.
The morphological and spectral parameters obtained for the HAWC sources using likelihood maximization can be found in Supplementary Table~\ref{tab:modellist}. HAWC observes LS~5039 as a point-like source with a significance of 11$\sigma$ 
above background and the best-fit coordinates of 
R.A. = 276.54$^{+0.004+0.003}_{-0.004-0.001}$ degrees, Decl. = -14.811$^{+0.02+0.002}_{-0.02-0.002}$ degrees.
The best-fit position from HAWC is within 0.03$^{\circ}$ of the optical measurement \cite{2004A&A...427..959C} and 0.04$^\circ$ from the radio source \cite{marti1998system}.


For the time-dependent analysis, 2,859~days of HAWC data are divided into shorter period maps. Using maps of single day transits, corresponding to $\sim$ 5 hours each or $\sim$ 0.05 in the orbital phase, no periodicity is observed due to limited statistics per map. Thus, we have produced two datasets matching to the inferior conjunction (INFC) of 0.45~$<$~$\phi$~$\leq$~0.9 and the superior conjunction (SUPC) of $\phi~\leq$~0.45 and $\phi$~ $>$0.9. The phase ranges correspond with the two phase ranges ($\phi$) of LS~5039 observed by H.E.S.S. \cite{hessls5039}. The INFC map contains 1,220~days of data and the SUPC map contains 1,504~days of data. Further information on the procedure of the time dependent analysis can be found in Methods~\ref{separatedata}.

\begin{figure}[ht!]
\begin{center}
\resizebox{1\textwidth}{!}{%
\includegraphics[width=0.5\textwidth]{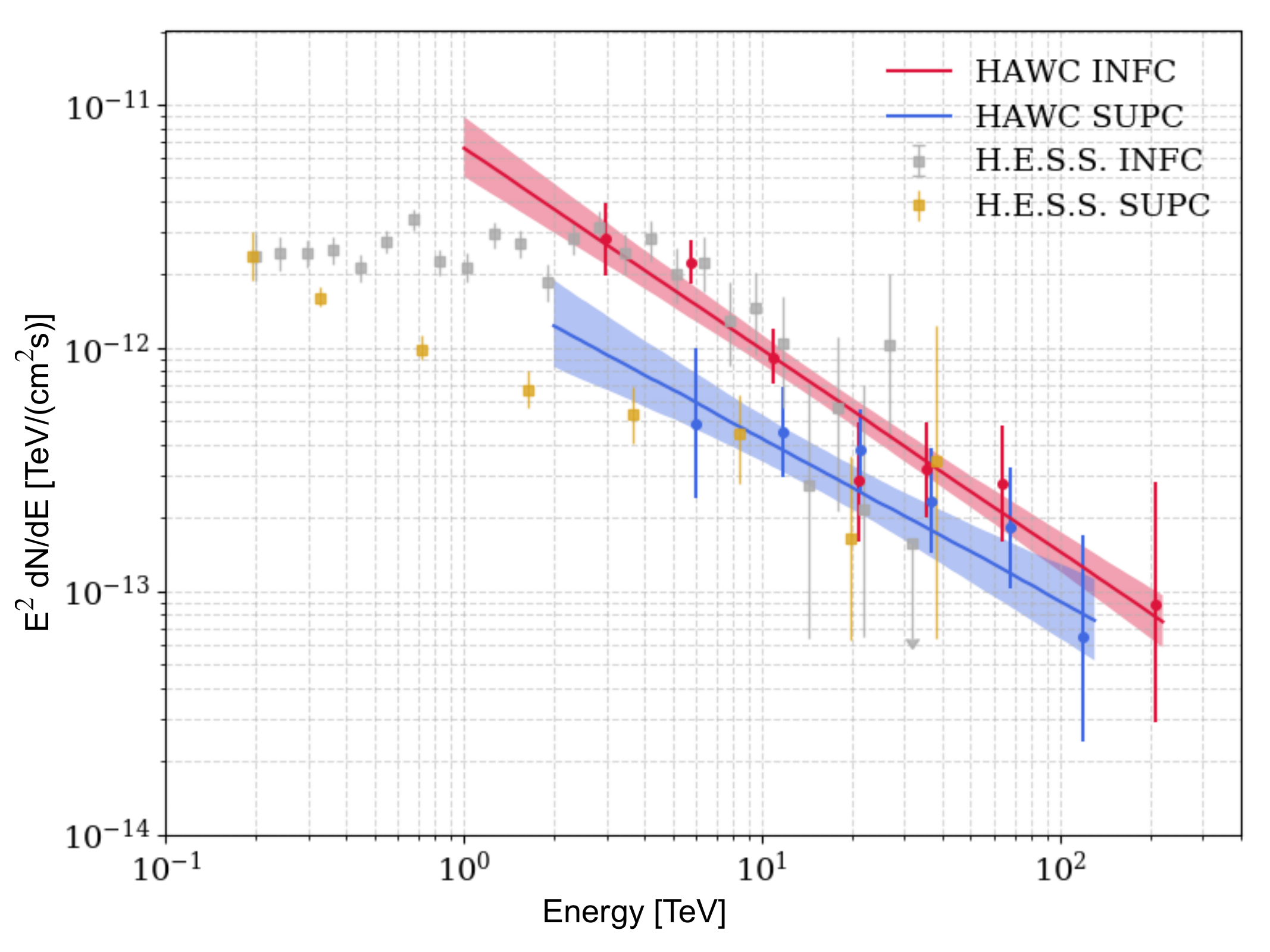}%
}
\caption{The spectral energy distribution~(SED) of LS 5039 at inferior conjunction~(red), and superior conjunction~(blue). The gray and yellow flux points are plotted from previous H.E.S.S. study~\cite{hessls5039}}
    
\label{fig:spectra}
\end{center}
\end{figure}   

We detect 9.8$\sigma$ excess from LS~5039 during INFC and 6.2$\sigma$ excess during SUPC after subtracting the background model (see Methods~\ref{sec:model}). Despite the longer phase interval for the SUPC than the INFC, LS 5039 is more significantly detected during INFC because of a higher flux. The LS~5039 spectral energy distribution (SED) plot in Figure~\ref{fig:spectra} shows the HAWC SUPC data points fitted with a simple power law spectrum~(Equation~\ref{power law}) in blue. This result is in agreement with the observational results from H.E.S.S. during SUPC in gold.
The HAWC INFC spectrum is best described by a power law that continues to $>100$~TeV without any signs of a cutoff, which is in disagreement with the cutoff power law spectrum observed by H.E.S.S.. The lower limit on the maximum energy measured by HAWC for LS~5039 is 208 TeV at 68\% confidence level for INFC. 
Upon examining other spectral models such as a power law with a cutoff or curved log-parabola spectrum, we have not found any improvements in the fit for the INFC and SUPC datasets.
The best-fit location and spectral parameters are listed in Table~\ref{tab:fit} for the fully time integrated case as well as for each of the phase sets. Between energies of 2 TeV and 118 TeV, the integrated flux from the INFC dataset differs from that of the SUPC dataset with a confidence level of 4.7$\sigma$ (see Methods~\ref{sigM}). This modulation extends beyond the previous H.E.S.S. results~\cite{hessls5039}, achieving a confidence level of 2.7$\sigma$ in the energy range of 40 TeV to 118 TeV. The modulation of the spectrum up to 100 TeV gives strong evidence of gamma-ray production inside the binary.

\begin{table}[ht!]
\caption{Fit parameters for each state } 
\label{tab:fit}
\begin{center}
\begin{tabular*}{\textwidth}{@{\extracolsep{\fill} } lcccc}
\hline
\hline
 &  Time Integrated & SUPC & INFC \\ 
 \hline
 $\text{R.A.}~ [^\circ$] & $276.54\pm 0.004~^{+0.003}_{-0.001}$ & $276.56\pm 0.04~^{+0.001}_{-0.000}$ & $276.52\pm 0.04~^{+0.003}_{-0.002}$ \\
 
 $\text{Decl.}~ [^\circ$] & $-14.81\pm 0.02~\pm 0.002$ & $-14.82\pm 0.04~^{+0.002}_{-0.001}$ & $-14.82\pm 0.027~\pm{0.002}$ \\
 
 $K$ [TeV$^{-1}$cm$^{-2}$s$^{-1}$] & 
 $\left( 1.54^{+0.20}_{-0.18}~^{+0.24}_{-0.22}\right) \times {10^{-15}}$ &
 $\left( 1.06^{+0.24~+0.04}_{-0.2~-0.14}\right) \times {10^{-15}}$ &
 $\left( 2.27^{+0.30~+0.14}_{-0.26~-0.29}\right) \times {10^{-15}}$ \\
 
 $\gamma$  & $2.76\pm 0.10~^{+0.03}_{-0.02}$ &
  $2.67\pm 0.16 ^{+0.03}_{-0.00}$ &
  $2.83\pm 0.09 ^{+0.02}_{-0.05}$ \\
 $E_{\text{min}}$  [TeV] &  0.4 & 2.1 & 1.0 \\
 $E_{\text{max}}$  [TeV] & 130 & 118 & 208 \\ 
 $\text{TS}$~(Equation~\ref{ts}) & 123 & 36 & 103 \\
 \hline
\end{tabular*}
\begin{tablenotes}
   \small
   \item $\textbf{Note}$: Spectral models are described in Equations~\ref{power law},~\ref{cutoff}. In the table, $K$ is the flux normalization at 16.8~TeV, $\gamma$ is the spectral index, and $E_{\text{min}}$ and $E_{\text{max}}$ represent the 68\%-confidence level upper limit on the minimum energy and lower limit on the maximum energy, respectively. The first uncertainty is statistical, and the second uncertainty is systematic~(see Methods~\ref{systematic}). 
 \end{tablenotes}
\end{center}
\end{table}

Ultra-relativistic electrons up-scattering ultra-violet photons emitted by the stellar companion are capable of producing the gamma rays observed by HAWC. The inverse Compton cross section depends on the product, $\xi$, of the energies of the interacting electrons and photons in units of $m_{\text{e}} \, c^2$. 
Since the hot stellar companion photons have temperatures of $k T \sim$ 3.3 eV, for multi-TeV energy electrons, $ \xi \gg 1$. Therefore, the inverse Compton scatterings should occur in the deep Klein-Nishina regime. In the Klein-Nishina regime, the LS~5039 spectrum observed by HAWC that extends to $>200$~ TeV requires electrons to be accelerated to at least 200 TeV by LS~5039 \cite{2004vhec.book.....A}.

As shown for a jet-like accelerator in the Materials~\ref{maxE}, the interplay of the competing acceleration and cooling mechanisms shapes the spectrum of the electrons depending on the physical conditions inside the accelerator. For LS~5039, a combination of dense radiation and high magnetic fields, and the limited orbital size of the binary system, make high-energy electron accelerations extremely challenging. As shown in Figure~\ref{fig:efficiency}, 200~TeV electrons, which are needed to explain HAWC emission, can be accelerated within the stellar photosphere only if $\eta \sim 1$, where $\eta$ is defined so that the acceleration rate is given by $\dot{\epsilon}=\eta \, qBc$ and ${E}_{\text{eff}} = \eta B$ is the projection of the electric field along the trajectory of a particle with electric charge $q$ inside a magnetic field of strength $B$. For example, in the case of Bohm diffusion, the acceleration efficiency is $\eta \sim 10 {(v/c)}^{-2}$. This suggests that the acceleration mechanism in LS 5039 is, for instance, more efficient than a diffusive shock acceleration in the Bohm regime. \cite{PhysRevD.66.023005}. Moreover, Figure~\ref{fig:efficiency}
indicates also that even for $\eta \sim 1$, $B$ cannot be significantly larger than 0.1~Gauss in order to prevent substantial synchrotron losses. For example, if $B \sim 1$ Gauss, then the synchrotron cooling time for 200 TeV electrons would be approximately $t_{\text{synch}} \sim  2 \left({\frac{B}{1 \text{Gauss}}}\right)^{-2} \, \left({\frac{E_\text{e}}{200 \text{TeV}}}\right)^{-1} {\rm s.}$ Furthermore, significantly lower values of $B$ can be ruled out to ensure efficient acceleration.

If the HAWC observation of the gamma-ray emission is of the hadronic origin, then this would unambiguously make LS~5039 an astronomical accelerator capable of producing PeV-scale hadrons. Protons may be boosted up to PeV energies, if the Larmor radius $r_\text{L} = \frac{E_\text{p}}{eB}$ is smaller than the acceleration region, $r$. This turns into a minimum maximum energy $E_{\text{max}} \lesssim 3 \left( \frac{r}{8 \times 10^8 {\rm cm}} \right) \left( \frac{B}{10^5 {\rm Gauss}}\right) $ PeV in the presence of magnetic fields of $ \sim 10^5$~Gauss within radii as small as 10$^{7}$-10$^{8}$~cm \cite{2006JPhCS..39..408A,doi:10.1146/annurev.aa.22.090184.002451,2024AdSpR..74.4276B}. 
Protons can efficiently produce gamma radiation through interactions with either dense materials or radiation fields. 
For example, protons colliding with dense materials provided by the winds of the stellar companion (up to $n_{\text{gas}} \sim 10^{10} \, \text{cm}^{-3}$ close to the binary \cite{Volkov_2021}) can produce gamma radiation. In another case, the radiation density inside and close to the binary system can exceed the gas density (see Materials for an estimate of the radiation density), causing protons with energies high enough to trigger photo-meson production by colliding with the stellar photons or the X-ray photons in the accretion disk \cite{2006JPhCS..39..408A}.
The timescale for proton-proton collisions and the subsequent pion decay is instantaneous and the proton cooling is given by $t_{\text{pp}} \sim \frac{1}{f_{\text{pp}} \sigma_{\text{pp}} n_{\text{gas}} c}$ s. With $\sigma_{\text{pp}} \sim$ 30 mbarn and the inelasticity f$_{\text{pp}} \sim 0.5 $, 
the cooling time of proton-proton collision is approximately 2.6 days, remarkably close to the binary period of 3.9 days.

In the following we assume a jet geometry, remarking that this does not imply automatically a micro-quasar scenario, as jet-like structures can arise also in wind binary systems \cite{10.1111/j.1365-2966.2008.13226.x}. Defining $k$ the proton acceleration efficiency in the jet, $L_p = k \,L_{jet}$, the required power of the jet to produce the HAWC radiation through proton-proton interactions is \cite{2006JPhCS..39..408A}
\begin{equation}
L_\text{{jet}} = 5 \times {10}^{36} \, {C(s)}^{-1/2}\, {\left(\frac{k}{0.1}\right)}^{-1/2} \, {\left(\frac{L_{\gamma}}{5.8 \times {10}^{33}}\right)}^{1/2} {\rm erg/s.}
\label{JetLum}
\end{equation}
The HAWC measured luminosity above 1 TeV is $L_{\gamma}= 5.8 \times {10}^{33}$ erg/s. 
The parameter $C(s)$ characterizes the geometry / gas density profile of the jet, $n_{\text{jet}} = {n_0(z/z_0)}^{-s}$ and $z_0 \sim 3 \times {10}^7$ cm. 
For s = 0 (cylindrical geometry), 2 (conical jet), and 1 (intermediate geometry), one has $C(s) = z_t/z_0$, $ln(z_t/z_0)$, and 1, respectively, with $z_t \sim 10^4 z_0$ the jet height where the protons escape the accelerating region \cite{2006JPhCS..39..408A}. Assuming Bondi-type accretion in LS 5039 a jet power of 10$^{36-37}$ erg/s can be obtained \cite{refId0}. In a colliding wind scenario, an upper limit on estimate for the jet luminosity is about $ 6 \times {10}^{36}$ erg/s \cite{Zabalza_2011}. Although it is not clear whether accretion or colliding winds is the source of power in LS 5039, in both scenarios, the binary jets can provide the required luminosity if $k \ge 0.1 $.

In the case of hadronic acceleration, the detected gamma rays should be accompanied by a flux of high energy neutrinos emerging from the decays of $\pi^{\pm}$ mesons produced by proton-proton and / or proton-$\gamma$ interactions. The flux of TeV neutrinos, which can be estimated on the basis of the detected TeV gamma-ray flux, taking into account the internal $\gamma \gamma \to e^{+} e^{-}$ absorption, depends significantly on the location of the gamma-ray production region. The minimum neutrino flux above 1~TeV is expected to be at the level of $10^{-12}~\text{TeV}\text{cm}^{-2}~\text{s}^{-1}$; while it can be up to 2 orders of magnitude higher \cite{2006JPhCS..39..408A}. The detectability of multi-TeV neutrinos depends heavily on the high energy cutoff in the spectrum of the parent protons. IceCube searches constrained neutrino flux at the HESS flux level \cite{2022ApJ...930L..24A}. If the spectrum of the accelerated protons from LS~5039 continues up to 1~PeV and beyond, the predicted neutrino fluxes can be probed by the planned Cubic Kilometer Neutrino Telescope, km3net \footnote{https://www.km3net.org/observatory}.

The observation by HAWC of the gamma-ray binary, LS~5039, shows that the spectrum, observed by H.E.S.S. up to several tens of TeV, extends beyond 100~TeV during its INFC and SUPC without a spectral cutoff. The detected modulation provides strong evidence that gamma rays up to very high energies are emitted within the orbit, which poses significant challenges to explaining the radiation with a leptonic scenario. The HAWC radiation can alternatively be interpreted as arising from collisions between PeV protons and either the dense gas or intense radiation regions within and close to the orbit of the binary. Follow-up observations from other experiments such as LHAASO or from future instruments in the Southern Hemisphere, such as The Southern Wide-field Gamma-ray Observatory (SWGO), at even higher energies could provide crucial evidence for proton-meson process within the high-density radiation regions. This would provide further insights into the acceleration mechanisms of the gamma-ray binary, LS~5039.

\newpage

{\bf References}

\newpage
\section{Methods}\label{sec:methods}

In this work, we have used 2,859 days of HAWC data that have been produced using the artificial neural network (NN) energy estimation method~\cite{abeysekara2019measurement}. This dataset contains improved angular resolution~\cite{albert2024performance} that allows us to study LS~5039 with minimum contamination. We have used the Multi-Mission Maximum Likelihood (3ML) framework~\cite{vianello2015multi} with HAWC plugins~\cite{abeysekara2022characterizing} to perform simultaneous multi-source likelihood modeling of the ROI.

\subsection{Modeling}\label{sec:model} A rectangular ROI shown in Figure~\ref{fig:map} is used to model and study VHE gamma-ray emissions from this region. We have applied a simple power law model to fit the spectra of sources at the pivot energy $E_0$:
\begin{equation}\label{power law}
    \frac{\text{d}N}{\text{d}E}=K \left(\frac{E}{E_{0}}\right) ^{-\gamma}.
\end{equation}

We have also used a power law with an exponential cutoff to test the potential curvature behavior on the spectrum of LS~5039:

\begin{equation}\label{cutoff}
    \frac{\text{d}N}{\text{d}E}=K \left(\frac{E}{E_{0}}\right)^{-\gamma}~\exp{\left(\frac{-E}{E_{\text{cut}}}\right)}.
\end{equation}
The cutoff energy, $E_{\text{cut}}$ is fitted as an additional free parameter for this spectral model.

We have assumed a 2 dimensional symmetric Gaussian morphological model for the extended sources in the multi-source fit. This model is presented below: 
\begin{equation}\label{extmodel}
    f(\vec{x}) = \left(\frac{180^\circ}{\pi}\right)^2 \frac{1}{2\pi \sqrt{\det{\Sigma}}}\, {\rm exp}\left( -\frac{1}{2} (\vec{x}-\vec{x}_0)^\intercal \cdot \Sigma^{-1}\cdot (\vec{x}-\vec{x}_0)\right),
\end{equation}
where $\overrightarrow{X_{0}}$ refers to the best-fit central location in equatorial coordinates, and $\Sigma$ is a covariance matrix defined as:
\begin{equation}
    \Sigma = U\Lambda U^\intercal,~ \\\Lambda = \left( \begin{array}{cc} \sigma^2 & 0 \\ 0 & \sigma^2 (1-e^2) \end{array}\right),~ \\ U = \left( \begin{array}{cc} \cos \theta & -\sin \theta \\ \sin \theta & \cos \theta \end{array}\right).
\end{equation}
Here, $\sigma$ refers to the standard deviation of the Gaussian distribution.

\subsection{Test Statistics}\label{sec:ts}
Likelihood ratio test statistic~(TS) is used to determine the significance of sources. 

\begin{equation}\label{ts}
    \text{TS} = 2\ln{\left(\frac{L_{\text{alt}}}{L_{\text{null}}}\right)},
\end{equation}
where $L_{\text{alt}}$ is the maximum likelihood from the alternative model, which contains background and target sources. $L_{\text{null}}$ is the likelihood for background sources only. According to the Wilks’ theorem~\cite{wilks1938large}, the TS is $\chi^{2}$ distributed, and thus the null model can be rejected with a probability from $\chi^{2}$ distribution. Thus, the $\sqrt{TS}$ can be used as a measure of significance when the degree of freedom equals one, which is expressible in terms of standard deviations $\sigma$.

\subsection{Systematic Source Search}\label{sec:sss}
The systematic source search method goes through the following procedure:
\begin{itemize}
    \item Firstly, the diffuse background emission (DBE) is fitted since our ROI is located close to the Galactic equator. We have used a simple DBE model centred around the Galactic latitude of $0^{\circ}$, a symmetric Gaussian profile along the Galactic latitude with a simple power law spectral model.
    For the fit, we have fixed the spectral index to -2.75~\cite{bloemen1989diffuse}.
    To cover the entire ROI, the DBE model spans between the galactic longitude of $15^{\circ}$ and $21^{\circ}$.
    
    \item Next, we consecutively add point sources to our nested model until $\Delta \text{TS}<16$ when compared to the previous model.

    \item The iterative extension test is performed starting with the most significant source found from the point source search. We accept the extended source model if:
    \begin{equation}
        \Delta \text{TS}_{\text{ext}}=2\ln{\frac{L_{\text{ext}}}{L_{\text{point}}}}>16.
    \end{equation} 

    During the extension test if replacing with an extended source results in the $\text{TS}$ of any nearby point sources to become $<16$, we accept these sources as being ``absorbed'' into the extended source and remove them from our nested model.

    \item The iterative curvature test is performed for the spectrum of each source. The power law with an exponential cutoff spectrum is fitted instead. We accept the new spectral model if:
    \begin{equation}
        \Delta \text{TS}_{\text{curvature}}=2\ln{\frac{L_{\text{cutoff}}}{L_{\text{powerlaw}}}}>16.
    \end{equation}
    
    \item Finally, we perform a likelihood fit with all the spatial and spectral parameters set free to obtain the final model for our ROI.
 
\end{itemize}
The best-fit model for this work is presented in Table~\ref{tab:modellist}.

\begin{table}[ht!]
\caption{Full model list} 
\label{tab:modellist}
\begin{center}
\def\arraystretch{1.5}
\begin{tabular*}{\textwidth}{@{\extracolsep{\fill} } lccccccc}
\hline
\hline
 Diffuse Background &  $\text{GLon.}~ [^\circ]$ & $\text{GLat.}~ [^\circ]$ & $K$ [$\text{TeV}^{-1}~\text{cm}^{-2}~\text{s}^{-1}$] &  $\gamma$ & \\ 
 \hline
 DBE & [15,21] & $\pm$ 1 & $\left( 2.7^{+0.5}_{-0.4}\right) \times {10^{-11}}$ & 2.75\\
 \hline
 \hline
 Background Sources &  $\text{R.A.}~ [^\circ]$ & $\text{Decl.}~ [^\circ]$ & $K$ [$\text{TeV}^{-1}~\text{cm}^{-2}~\text{s}^{-1}$] &  $\gamma$ & Cutoff~(TeV) & Ext.~ [$^\circ$] \\
 \hline
 HAWC J1825-136 & $276.49^{+0.01}_{-0.01} $  & $-13.63^{+0.012}_{-0.012}$ & $\left( 1.23^{+0.11}_{-0.12} \right) \times {10^{-13}}$ & $2.03^{+0.08}_{-0.08}$ & $85^{+16}_{-14}$ & $0.23^{+0.011}_{-0.011}$ \\
 
 HAWC J1825-130 & $276.48^{+0.012}_{-0.012} $  & $-12.97^{+0.018}_{-0.018}$ & $\left( 4.3^{+0.4}_{-0.4} \right) \times {10^{-14}}$ & $2.38^{+0.04}_{-0.04}$ &  & $0.133^{+0.013}_{-0.013}$ \\
 
 HAWC J1825-125  & $276.49^{+0.04}_{-0.04} $  & $-12.52^{+0.06}_{-0.06}$ & $\left( 6.7^{+0.22}_{-0.16} \right) \times {10^{-15}}$ & $2.86^{+0.24}_{-0.24}$ & &  \\
 
HAWC J1825-140  & $276.31^{+0.04}_{-0.04} $  & $-14.05^{+0.06}_{-0.06}$ & $\left( 3.19^{+0.32}_{-0.35} \right) \times {10^{-13}}$ & $2.42^{+0.13}_{-0.13}$ & $31^{+7}_{-5}$ & $0.72^{+0.05}_{-0.05}$ \\
 
 \hline
 \hline
 LS5039 &  $\text{R.A.}~ [^\circ]$ & $\text{Decl.}~ [^\circ]$ & $K$ [$\text{TeV}^{-1}~\text{cm}^{-2}~\text{s}^{-1}$] &  $\gamma$ & Cutoff~(TeV) & Ext.~ [$^\circ$] \\
 \hline
 HAWC J1826-148 & $276.54^{+0.004}_{-0.004} $  & $-14.81^{+0.02}_{-0.02}$ & $\left( 1.54^{+0.18}_{-0.20} \right) \times {10^{-15}}$ & $2.76^{+0.10}_{-0.10}$ &  &  \\
 \hline
 \end{tabular*}
 \begin{tablenotes}
   \small
   \item $\textbf{Note}$:~DBE only fits the flux, with the index fixed at 2.75. The pivot energy for background sources is set at $E_{0}$ = 7~TeV. For LS 5039 $E_{0}$ = 16.8~TeV, which is optimized by minimizing the correlation between spectral index and flux. The uncertainty in this table represents the statistical error only.
 \end{tablenotes}
\end{center}
\end{table}

\subsection{Dividing HAWC Data for SUPC and INFC}\label{separatedata}

We have used T$_{0}$ = HJD~2451943.09 from \cite{2005MNRAS.364..899C} and consecutively added P$_{\text{orb}}$ = 3.90603 days until it reached the start of HAWC data. Then, we have made the INFC and SUPC maps by stacking the data corresponding to the known phases, namely, 0.45~$<$~$\phi$~$\leq$~0.9 for INFC and $\phi~\leq$~0.45 and $\phi$~ $>$0.9 for SUPC. There are $\sim$ 2,859 full transits across the HAWC sky used for the analysis in total. 1220 transits belong to INFC and 1504 belong to SUPC. Note that we have removed overlapping data containing both INFC and SUPC periods~($\sim$5\% of the total data volume). 
\begin{figure}[ht!]
\begin{center}
\resizebox{1.0\textwidth}{!}{%
\includegraphics[width=0.5\textwidth]{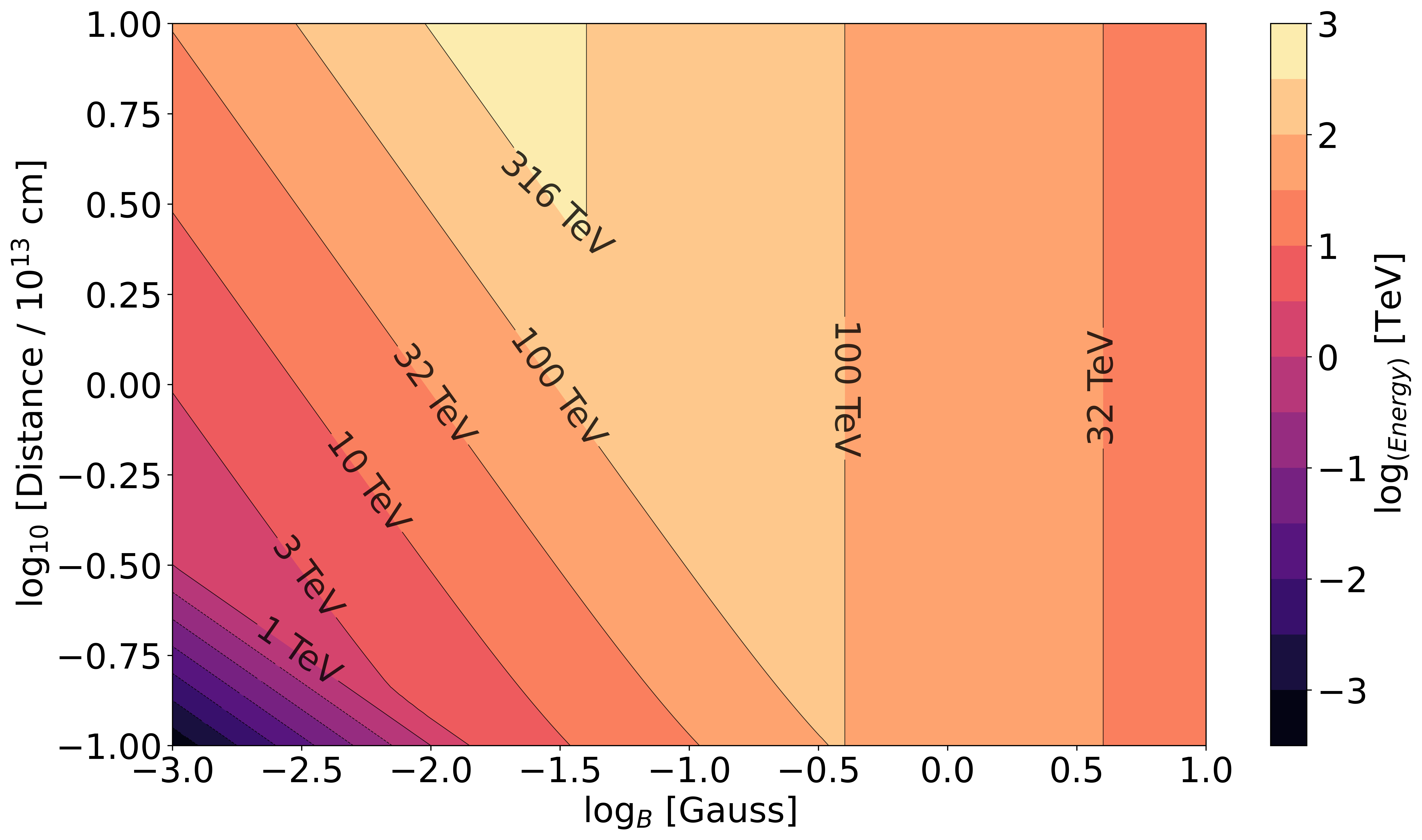}%
\quad
\includegraphics[width=0.5\textwidth]{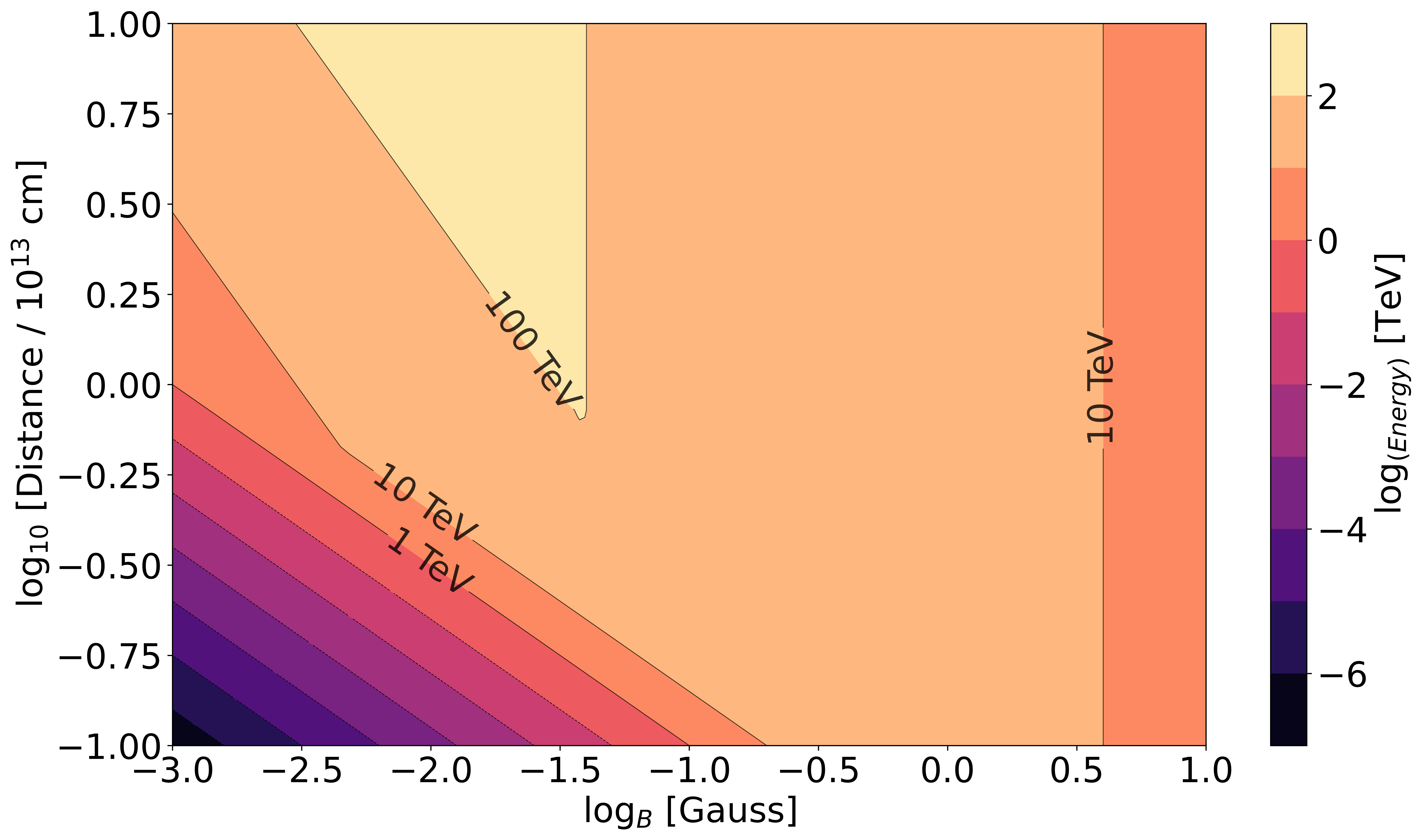}%
}
\caption{Maximum electron energy achievable within LS 5039 as a function of the acceleration efficiency (left $\eta$=1, right $\eta$=10) and of the magnetic field, B in Gauss (x-axis), and of the distance from the companion star, $d_{13}=\frac{d}{{10}^{13} \, {cm}}$ (y-axis).} 
    
\label{fig:efficiency}
\end{center}
\end{figure}   

\subsection{Maximum electron energy}\label{maxE}

We here estimate the maximum energy in electrons which can be achieved in LS 5039 by comparing the competing acceleration and cooling mechanisms at work in the source. We assume that the accelerator has a jet-like structure.

The companion star in LS~5039 emits a black-body radiation with temperature $kT\sim 3.3$~eV and has luminosity of $L_{\text{star}} = 7 \times {10}^{38}$~erg/s. We assume that the energy density of the stellar radiation, $w$, decreases as a function of distance, $d = \sqrt{({R_{\text{orb}}}^2 + z^2)}$, from the base of the jet as $w = \frac{L_{\text{star}}}{4 c \pi {(R_{\text{orb}}}^2 + z^2)}$, where $R_{\text{orb}} \sim {10}^{12}$~cm and $z$ is the height from the orbit plane. $w$ changes from $\sim$ 1000 erg/cm$^3$ at the base of the jet to $\sim$ 10 erg/cm$^3$ at a distance $z \sim 10 \, R_{\text{orb}}$ or $\sim$ 0.1 erg/cm$^3$ at a distance $z \sim 100 \, R_{\text{orb}}$. In the deep Klein Nishina regime, the cooling time is dependent on the distance as 
\begin{equation}
t_{\text{KN}} \sim  2.5 \times {10}^4 \, ({d_{13}})^{2} \, \left({\frac{E_\text{e}}{100 \text{TeV}}}\right)^{0.7}~{\rm s,}
\label{tKN}
\end{equation}
where $d_{13} = \sqrt{z^2 + R_{\text{orb}}^2}/{10}^{13}$ cm, and $E_{\text{e}}$ is the electron energy \cite{2006JPhCS..39..408A}.
With acceleration environments requiring high $B$-fields, the synchrotron cooling time is generally shorter than the KN time,
\begin{equation}
t_{\text{synch}} \sim  4 \left({\frac{B}{1 \text{Gauss}}}\right)^{-2} \, \left({\frac{E_\text{e}}{100 \text{TeV}}}\right)^{-1} {\rm s.}
\label{tKN}
\end{equation}
The acceleration time is given by 
\begin{equation}
t_{\text{acc}}= \eta r_\text{L}/c \sim 10 \eta \left({\frac{E_\text{e}}{100 \text{TeV}}}\right) \left({\frac{B}{1 \text{Gauss}}}\right)^{-1} {\rm s,}
\label{tacc}
\end{equation}
with $r_\text{L} = \frac{E_\text{e}}{eB}$ and {$\eta$ the acceleration efficiency previously defined}.
Following \cite{2008MNRAS.383..467K}, the maximum electron energy is achieved when $t_{\text{acc}} = t_{\text{cool}} = \text{min} [t_{\text{KN}}, t_{\text{synch}}]$ where $t_{\text{cool}}$ is the Klein-Nishina / synchrotron cooling timescale and $t_{\text{acc}}$ is the acceleration timescale. Equating $t_{\text{acc}}=t_{\text{KN}}$, we obtain the constraint:
\begin{equation}
E_{\text{max}} <   2 \times {10}^{13} \, [{{\eta}^{-1} \, \left(\frac{B}{1 \text{Gauss}}\right) \, d_{13}^2 ]}^{1/0.3} \, {\rm TeV}.
\label{emaxkn}
\end{equation}
Alternatively, equating $t_{\text{acc}}=t_{\text{synch}}$, we obtain the constraint:
\begin{equation}
E_{\text{max}} <  60 \, {\eta}^{-1/2} \, {\left(\frac{B}{1 \text{Gauss}}\right)}^{-1/2} \, {\rm TeV}.
\label{emaxsyn}
\end{equation}
Finally, the Larmor radius of the particle has to be smaller than the size of the accelerator or at most $z$, namely, $r_\text{L} = 3.3 \times {10}^{10} {(\frac{B}{\text{Gauss}})}^{-1} \, (\frac{E}{1 \text{TeV}})
< z$, thus,
\begin{equation}
E_{\text{max}} <  300 \, \left({\frac{B}{1 \text{Gauss}}}\right) \frac{\sqrt{(d_{13}^2 {10}^{26} - R_{\text{orb}}^2)}}{10^{13}} \, {\rm TeV}.
\label{emaxrad}
\end{equation}
Formally, if $B \ge$ 1 G then electrons can be accelerated to hundred TeVs. Nonetheless, if $B \ge$ 1 G, synchrotron cooling dominates over KN cooling ($t_{\text{synch}} < t_{\text{KN}}$). Combining the previous conditions \ref{emaxkn},\ref{emaxsyn},\ref{emaxrad} produce the constraints on the phase space in $d_{13}$ ($y$-axis) and B ($x$-axis) shown in Fig~\ref{fig:efficiency}. In Fig.~\ref{fig:efficiency} electrons can be accelerated up to hundreds TeV within the binary region only if $\eta\sim 1$ and B $ll$ 1 Gauss. We finally note if the highest energy photons above several tens of TeV were not unmodulated, then they could be emitted by electrons accelerated up to > 100 TeV in regions outside the binary ($d_{13} \gg 1)$ where $B \gg $ 1 G.

\subsection{Significance of Flux Modulation}\label{sigM}

To robustly estimate the statistical significance of modulation between INFC and SUPC, we have performed separate likelihood fits to determine the best-fit flux normalization parameters from a power-law model~(see Equation~\ref{power law}), represented as $K_{\text{INFC}}$ and $K_\text{{SUPC}}$. Subsequently, we have selected an energy range with an upper limit of minimum energy of 2 TeV and a lower limit of maximum energy of 118 TeV from SUPC. The integrated fluxes,  $F_{\text{INFC}}$ and $F_{\text{SUPC}}$, are calculated within this energy range. 

To account for uncertainties in the flux normalization, we propagated errors for the integrated flux F using the following equation:
\begin{equation}
    \sigma_{\text{i}} = F \cdot \frac{\sigma_{{\text{k}}}}{K}
\end{equation}
Here, $\sigma_{\text{i}}$ is the error in the integrated flux, F is the integrated flux, $\sigma_{\text{k}}$ is the error, and K is the flux normalization from the likelihood fit.

The significance of the modulation denoted as $\sigma_{\text{M}}$, is then quantitatively determined based on the formulation:
\begin{equation}\label{sig-modulation}
    \sigma_{\text{M}}=\frac{|F_{\text{INFC}}-F_{\text{SUPC}}|}{\sqrt{\sigma^{2}_{\text{INFC}}+\sigma^{2}_{\text{SUPC}}}}.
\end{equation}
This analysis has yielded a modulation significance of 4.7 $\sigma$ within the energy range of 2 TeV to 118 TeV. The corresponding errors, $\sigma_{\text{INFC}}$ and $\sigma_{\text{SUPC}}$, encapsulate the uncertainties propagated from errors obtained through the likelihood fit. For asymmetric errors, we have averaged the errors before performing error propagation. Using the same method, we have obtained a modulation significance of 2.7 $\sigma$ within the energy range of 40 TeV to 118 TeV.

\subsection{Systematic Analysis}\label{systematic}

The systematic uncertainty presented in Table \ref{tab:fit} has been calculated by performing likelihood fits of our model with different detector response functions. Five major contributions of systematic errors have been investigated, including charge uncertainty, energy estimator, PMT absolute quantum efficiency, time dependence, and late light effects \cite{abeysekara2017observation}.

The final systematic error is computed by summing the components in quadrature. Each of the systematic components and the total systematic uncertainties are presented in Table~\ref{tab:systematictable}.

\begin{table}[h!]
\centering
\caption{Systematic Uncertainties}
\label{tab:systematictable}
\begin{tabular}{lcc|cc|cc}
\hline
\hline
& \multicolumn{2}{c}{Average State} & \multicolumn{2}{c}{INFC} & \multicolumn{2}{c}{SUPC} \\
\hline
Uncertainty Origin & $K$  & $\gamma$ &  $K$  & $\gamma$ & $K$  & $\gamma$ \\
& {[}TeV$^{-1}$ \text{cm}$^{-2}$ s$^{-1}${]} $\times 10^{-16}$ &  & {[}TeV$^{-1}$ \text{cm}$^{-2}$ s$^{-1}${]} $\times 10^{-16}$ &  & {[}TeV$^{-1}$ \text{cm}$^{-2}$ s$^{-1}${]} $\times 10^{-16}$ & \\
\hline
Late light & $+0.2 $ & $+0.01$ &  $+0.7   $ & $+0.01$  & $+0.3  $ & $+0.01$ \\
& $-1.9 $ & $-0.02$  & $-2.6 $ & $-0.04$  & $-1.3 $ & $-0.00$ \\
Charge uncertainty & $+0.4 $ & $+0.02$ &  $+1.0  $ & $+0.01$  & $+0.3 $ & $+0.02$ \\
& $-0.8 $ & $-0.01$  & $-0.9 $ & $-0.03$  & $-0.5 $ & $-0.00$ \\
Time dependence & $+0.1  $ & $+0.01$ &  $+0.7  $ & $+0.01$  & $+0.1  $ & $+0.02$ \\
& $-0.1 $ & $-0.00$  & $-0.00 $ & $-0.01$  & $-0.1 $ & $-0.00$ \\
PMT threshold & $+0.00 $ & $+0.00$  & $+0.1 $ & $+0.00$ & $+0.00 $ & $+0.01$ \\
 & $-0.7 $ & $-0.00$  & $-0.8 $ & $-0.01$ & $-0.3 $ & $-0.00$ \\
Energy estimator & $+2.4  $ & $+0.02$  &  &  &  &  \\
& $-0.00 $ & $-0.00$  &  &  &  &  \\
 \hline
Quadratic sum & $+2.44 $ & $+0.032$ &  $+1.41  $ & $+0.017$  & $+0.44  $ & $+0.032$ \\
& $-2.18 $ & $-0.022$  & $-2.87 $ & $-0.052$  & $-1.43 $ & $-0.000$ \\
\hline
\end{tabular}
\end{table}

\newpage
{\bf References}
\vspace{1em}

\makeatletter
\renewcommand{\thebibliography}[1]{%
  \section*{\refname}%
  \list{\@biblabel{\arabic{enumi}}}{%
    \settowidth\labelwidth{\@biblabel{#1}}%
    \leftmargin\labelwidth
    \advance\leftmargin\labelsep
    \usecounter{enumi}%
    \let\p@enumi\@empty
    \renewcommand\theenumi{\arabic{enumi}}%
  }%
  \sloppy
  \clubpenalty4000
  \@clubpenalty \clubpenalty
  \widowpenalty4000%
  \sfcode`\.\@m
}
\makeatother

\setcounter{enumi}{26} 

\end{document}